\documentclass[aps,preprint]{revtex4}
\usepackage[dvips]{graphicx}
\begin{document}
\draft
\title{\bf Investigation of nodal domains in the chaotic microwave ray-splitting rough billiard}
\author{Oleh Hul, Nazar Savytskyy, Oleg Tymoshchuk, Szymon Bauch and Leszek Sirko}

\address{Institute of Physics, Polish Academy of Sciences, Aleja  Lotnik\'{o}w 32/46, 02-668 Warszawa, Poland}
\date{\today}
\date{October 20, 2005}

\bigskip

\begin{abstract}

We study experimentally nodal domains of wave functions (electric
field distributions) lying in the regime of Shnirelman ergodicity
in the chaotic  microwave half-circular ray-splitting rough
billiard. For this aim the wave functions $\Psi_N$  of the
billiard were measured up to the level number $N=415$. We show
that in the regime of Shnirelman ergodicity ($N>208$) wave
functions of the chaotic half-circular microwave ray-splitting
rough billiard are extended over the whole energy surface and the
amplitude distributions are Gaussian. For such ergodic wave
functions the dependence of the number of nodal domains $\aleph_N$
on the level number $N$ was found. We show that in the limit $N
\rightarrow \infty $ the least squares fit of the experimental
data yields $\aleph_N/N \simeq 0.063 \pm 0.023$ that is close to
the theoretical prediction $\aleph_N/N \simeq 0.062$. We
demonstrate that for higher level numbers $N \simeq 215-415$ the
variance of the mean number of nodal domains $\sigma^2_N /N$ is
scattered around the theoretical limit $\sigma^2_N /N \simeq
0.05$. We also found that the distribution of the areas $s$ of
nodal domains has power behavior $n_s \propto s^{-\tau}$, where
the scaling exponent is equal to $\tau = 2.14 \pm 0.12$. This
result is in a good agreement with the prediction of percolation
theory.

\end{abstract}


\bigskip
\maketitle

\smallskip

In recent theoretical papers by  Bogomolny and Schmit
\cite{Bogomolny2002} and Blum {\it et al.} \cite{Blum2002} the
distributions of the nodal domains of real wave functions
$\Psi(x,y)$ in 2D quantum systems (billiards) have been
considered.  Nodal domains are regions where a wave function
$\Psi(x,y)$ has a definite sign. The condition $\Psi(x,y)=0$
determines a set of nodal lines which separate nodal domains.
Bogomolny and Schmit \cite{Bogomolny2002} have proposed a very
fruitful, percolationlike, model for description of properties of
the nodal domains of generic chaotic system. Using this model they
have shown that the distribution of nodal domains of quantum wave
functions of chaotic systems is universal. Blum {\it et al.}
\cite{Blum2002} have shown that the systems with integrable and
chaotic underlying classical dynamics can be distinguished by
different  distributions of the number of nodal domains. In this
way they provided a new criterion of quantum chaos, which is not
directly related to spectral statistics.

Theoretical findings of Bogomolny and Schmit \cite{Bogomolny2002}
and  Blum {\it et al.} \cite{Blum2002} have been recently tested
in the experiment with the microwave half-circular rough billiard
by Savytskyy {\it et al.} \cite{Savytskyy2004}.

In this paper we present the first experimental investigation of
nodal domains of wave functions of the chaotic microwave
ray-splitting rough billiard. Ray-splitting systems are a new
class of chaotic systems in which the underlying classical
mechanics is non-Newtonian and non-deterministic
\cite{BLUM96,SIR97,BLUMEL2001}. In ray-splitting systems a wave
which encounters a discontinuity in the propagation medium splits
into two or more rays travelling usually away from the
discontinuity. Ray splitting occurs in many fields of physics,
whenever the wave length is large in comparison with the range
over which the potential changes.  Ideal model systems for the
investigation of ray-splitting phenomena are ray-splitting
billiards \cite{COUC92,BLUMEL2001} and microwave cavities with
dielectric inserts \cite{SIR97,BAUCH98,HLUSH2000,Savytskyy2001}.
Measurements of  wave functions of ray-splitting systems are very
demanding because  in principle they require the direct access to
the all parts of the system \cite{Stoeckmann2001} including those
filled with ray-splitting media, such as dielectric in the case of
ray-splitting microwave billiards. This is one of the main reasons
for which only low wave functions ($N\leq 100$) of ray-splitting
billiards have been measured so far \cite{Stoeckmann2001}. In this
paper we use a new method of the reconstruction of wave functions
introduced by Savytskyy and Sirko \cite{Savytskyy2002} which in
the case  of the half-circular microwave ray-splitting rough
billiard allowed for the reconstruction of wave functions with the
level numbers $N\leq 415$.

\begin{figure}[!]
\begin{center}
\rotatebox{270} {\includegraphics[width=0.5\textwidth,
height=0.6\textheight, keepaspectratio]{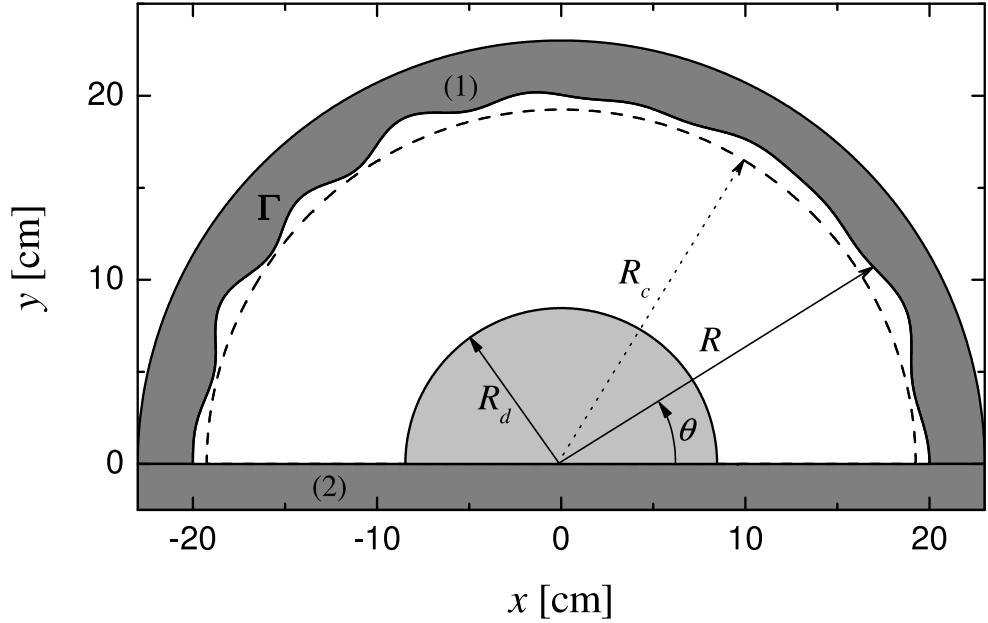}} \caption{Sketch of
the chaotic half-circular microwave ray-splitting rough billiard
which consists a half-circular Teflon insert of radius $R_d=8.465$
cm. Dimensions are given in cm. The cavity sidewalls are marked by
1 and 2 (see text). Squared wave functions $|\Psi_N(R_c,\theta
)|^2$ were evaluated on a half-circle of  fixed radius $R_c=19.25$
cm. Billiard's rough boundary is marked by  $\Gamma
$.}\label{Fig1}
\end{center}
\end{figure}

In the experiment we used the thin (height $h=8$ mm) aluminium
cavity in the shape of a rough half-circle (Fig. \ref{Fig1}) which
consisted a half-circular Teflon insert of radius $R_d=8.465$ cm.
The insert had the same height as the rough cavity. The microwave
cavity simulates the rough ray-splitting quantum billiard due to
the equivalence between the Schr\"odinger equation and the
Helmholtz equation \cite{BLUMEL2001}. This equivalence remains
valid for frequencies less than the cut-off frequency $\nu_c
=c/2\eta h \simeq 13.1$ GHz, where c is the speed of light and
$\eta=1.425$ is the index of refraction of the Teflon insert.

The cavity sidewalls were made of two segments. The rough segment
1 is described by the radius function
$R(\theta)=R_{0}+\sum_{m=2}^{M}{a_{m}\sin(m\theta+\phi_{m})}$,
where  the mean radius $R_0$=20.0 cm, $M=20$, $a_{m}$ and
$\phi_{m}$ are uniformly distributed on [0.084,0.091] cm and
[0,2$\pi$], respectively, and $0\leq\theta<{\pi}$.  It is
important to note that  we used a rough half-circular cavity
instead of a rough circular cavity because in this way we avoided
nearly degenerate low-level eigenvalues
\cite{Hlushchuk01b,Hlushchuk01}. Additionally, a half-circular
geometry of the cavity was necessary for the accurate measurements
of the electric field distributions inside the billiard.

According to \cite{Frahm97} the roughness of a  billiard may be
characterized by the function $k(\theta)=(dR/d\theta)/R_0$. The
roughness parameter $\tilde k$ defined as  the angle average of
the function $k(\theta)$ was for our billiard $\tilde k=(\left
<k^{2}(\theta)\right >_{\theta})^{1/2}\simeq  0.200$. In such a
billiard the dynamics is diffusive in orbital momentum due to
collisions with the rough boundary because the roughness parameter
$\tilde k $ is much larger the chaos border parameter
$k_c=M^{-5/2}=0.00056$ \cite{Frahm97}. The roughness parameter
$\tilde k $ determines also other properties of the billiard
\cite{Frahm}. The eigenstates are localized for the level number
$N < N_e = 1/128 \tilde k^4=5$. The border of Breit-Wigner regime
is given by $N_W = M^2/48\tilde k^2 \simeq 208$. It means that
between $N_e < N < N_W$ Wigner ergodicity \cite{Frahm} ought to be
observed and for $N > N_W$ Shnirelman ergodicity should emerge. In
the regime of Shnirelman ergodicity wave functions have to be
uniformly spread out in the billiard \cite{Shnirelman}. In this
paper we focus our attention on Shnirelman ergodicity regime.

It is worth noting that rough billiards and related systems are of
considerable interest elsewhere, e.g. in the context of  microdisc
lasers \cite{Yamamoto,Stone}, light scattering in optical fibers
\cite{Doya2002}, ballistic electron transport in microstructures
\cite{Blanter}, dynamic localization \cite{Sirko00} and
localization in discontinuous quantum systems \cite{Borgonovi}.

In order to measure the wave functions (electric field
distributions inside the microwave billiard), which are
indispensable in investigation of nodal domains, we used a new,
very effective method described in \cite{Savytskyy2002}. It is
based on the perturbation technique and construction of the
``trial functions".

Following \cite{Savytskyy2002} we will show that the  wave
functions $\Psi_N(r,\theta )$ (electric field distribution
$E_N(r,\theta )$ inside the cavity)  of the billiard can be
determined from the form  of electric field $E_N(R_c, \theta )$
evaluated  on a half-circle of  fixed radius $R_c$ (see
Fig.~\ref{Fig1}).

The first step in evaluation of $E_N(R_c, \theta )$ is measurement
of $|E_N(R_c, \theta )|^2$. For this purpose the perturbation
technique developed in \cite{Slater52} and used successfully in
\cite{Slater52,Sridhar91,Richter00,Anlage98} was applied. In this
method a small perturber is introduced inside the cavity to alter
its resonant frequency according to
$$\nu -\nu_N =\nu_N(aB_N^2-bE_N^2), \eqno(1)$$ where $\nu_N $ is the
$N$th resonant frequency of the unperturbed cavity, $a$ and $b$
are geometrical factors.  Equation (1)  can be used to evaluate
$E_N^2$ only when the term containing magnetic field $B_N$ is
sufficiently small. In order to minimize the influence of $B_N$ on
the frequency shift $\nu -\nu_N $ a small piece of a metallic pin
(3.0 mm in length and 0.25 mm in diameter) was used as a
perturber. The perturber was attached to the micro filament line
hidden in the groove (0.4 mm wide, 1.0 mm deep) made in the
cavity's bottom wall along the half-circle $R_c$ and moved by the
stepper motor. Application of such a small pin perturber reduced
the largest positive frequency shifts to the uncertainty of
frequency shift measurements (15 kHz). It was verified that the
presence of the narrow groove in the bottom wall of the cavity
caused only very small changes $\delta \nu_N$ of the
eigenfrequencies $\nu_N$ of the cavity $|\delta \nu_N|/\nu_N \leq
10^{-4}$. Therefore, its influence into the structure of the
cavity's wave functions was also negligible. A big advantage of
using  the perturber that was attached to the line, was connected
with the fact that the perturber was always vertically positioned,
which is crucial in the measurements of the square of electric
field $E_N$. The influence of the thermal expansion of the Teflon
insert and the aluminium cavity into its resonant frequencies was
eliminated by stabilization of the temperature of the cavity with
the accuracy of $0.05^{\circ} $.

\begin{figure}[!]
\begin{center}
\rotatebox{0} {\includegraphics[width=0.5\textwidth,
height=0.8\textheight, keepaspectratio]{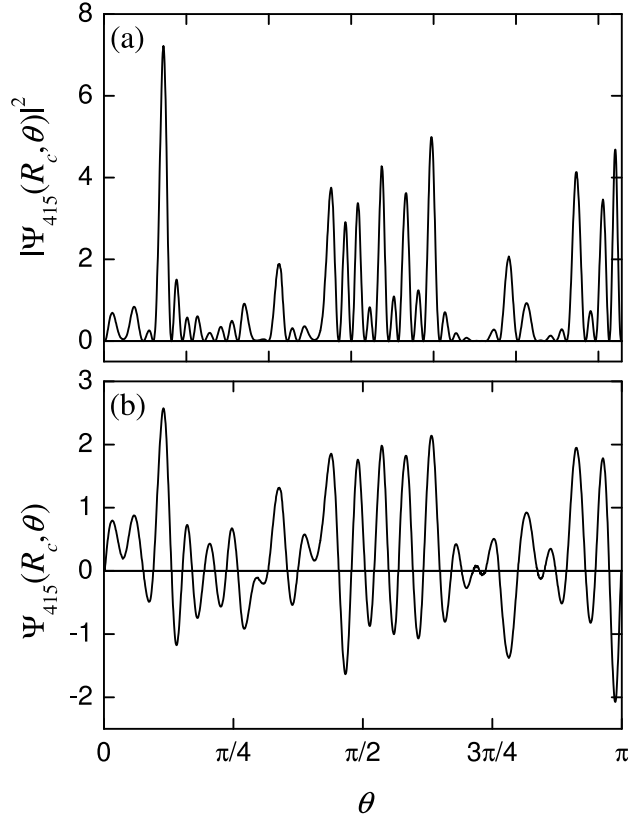}} \caption{Panel (a):
Squared wave function $|\Psi_{415}(R_c,\theta )|^2$ (in arbitrary
units) measured on a half-circle with radius $R_c=19.25$ cm
($\nu_{415} \simeq 12.98$ GHz). Panel (b): The ``trial wave
function" $\Psi_{415}(R_c,\theta )$ (in arbitrary units) with the
correctly assigned  signs, which was used in the reconstruction of
the wave function $\Psi_{415}(r, \theta )$ of the billiard (see
Fig. \ref{Fig3}). }\label{Fig2}
\end{center}
\end{figure}

The regime of Shnirelman ergodicity for the experimental rough
billiard is defined for $N > 208$. Using a field perturbation
technique we measured squared wave functions $|\Psi_N(R_c,\theta
)|^2$  for 30 modes within the region $215\leq N \leq 415$. The
range of corresponding eigenfrequencies was from $\nu_{215} \simeq
9.42$ GHz to $\nu_{415} \simeq 12.98$ GHz. The measurements were
performed at 0.36 mm steps along a half-circle with fixed radius
$R_c=19.25$ cm. This step was small enough to reveal in details
the space structure of high-lying levels. In Fig. \ref{Fig2} (a)
we show the example of the squared wave function $|\Psi_N (R_c,
\theta )|^2$ evaluated for the level number $N=415$. The
perturbation method used in our measurements allows us to extract
information about the wave function amplitude $|\Psi_N(R_c, \theta
)|$ at any given point of the cavity but it doesn't allow to
determine the sign of $\Psi_N(R_c, \theta )$ \cite{Stein95}.
However, the determination of the sign of the wave function
$\Psi_N(R_c, \theta )$ is crucial in the procedure of the
reconstruction of the full wave function $\Psi_N(r, \theta )$ of
the billiard. The papers \cite{Savytskyy2002,Savytskyy2004}
suggest the following sign-assignment strategy. First one should
identify of  all close to zero minima of  $|\Psi_N(R_c, \theta
)|$. Then the sign ``minus" is arbitrarily assigned to the region
between the first and the second minimum, ``plus" to the region
between the second minimum and the third one and so on. In this
way the ``trial wave function" $\Psi_N(R_c, \theta )$ is
constructed. If the assignment of the signs is correct the wave
function $\Psi_N(r, \theta )$ should be reconstructed inside the
billiard with the boundary condition $\Psi_N(r_{\Gamma },
\theta_{\Gamma } )=0$.

The wave function of a rough ray-splitting half-circular billiard
outside of the half-circular Teflon insert ($r\geq R_d$) may be
expanded in terms of  Hankel functions
$$
\Psi^{out}_N(r, \theta ) = \sum_{s=1}^L a_s \Omega_s
(k_Nr)\sin(s\theta ),  \eqno(2)
$$
where $ \Omega_s
(x)=Re\{H^{(2)}_{s}(x)+S_{ss}(k_NR_d)H^{(1)}_{s}(x)\} $ and
$k_N=2\pi \nu_N /c$. $H^{(1)}_{s}(x)$ and $H^{(2)}_{s}(x)$ are
Hankel functions of the first and the second kind, respectively.
The matrix $S_{ss'}(k_N R_d)$ is defined as follows
\cite{Hentschel2002}
$$
S_{ss'}(k_N R_d)=-\frac{H^{(2)'}_{s}(k_N R_d)-\eta [J'_s(\eta k_N
R_d)/J_s(\eta k_N R_d)]H^{(2)}_{s}(k_N R_d)} {H^{(1)'}_{s} (k_N
R_d)-\eta [J'_s(\eta k_N R_d)/J_s(\eta k_N R_d)]H^{(1)}_{s}(k_N
R_d)}\delta_{ss'}, \eqno(3)
$$
where the derivatives of  Hankel and Bessel functions are marked
by primes. In Eq. (2) the number of basis functions is limited to
$L=k_N r_{max} + 3$, where $r_{max}=20.7$ cm is the maximum radius
of the cavity. $l_N^{max} = k_N r_{max}$ is a semiclassical
estimate for the maximum possible angular momentum for a given
$k_N$. The functions with angular momentum $s > l_N^{max}$
describe evanescent waves. We checked  that the basis of $L$ wave
functions was large enough to properly reconstruct billiard's wave
functions. The coefficients $a_s$ may be determined from the
``trial wave functions" $\Psi_N(R_c, \theta )$ via
$$
a_s=[\frac{\pi}{2}\Omega_s (k_N
R_c)]^{-1}\int_0^{\pi}\Psi_N(R_c,\theta)sin(s\theta)d\theta.
\eqno(4)
$$

The wave functions of the billiard inside the Teflon insert
($r\leq R_d$) may be expanded in terms of circular waves
$$
\Psi^{in}_N(r, \theta ) = \sum_{s=1}^{L'} a^{'}_s J_{s}(\eta
k_Nr)\sin(s\theta ). \eqno(5)
$$

In Eq.~(5) the number of basis functions was limited to $L'=\eta
k_N R_d$. The  coefficients  $a_s$  given by Eq.~(4) and the
continuity condition fulfilled at the border of the dielectric
insert $\Psi^{out}_N(R_d, \theta )= \Psi^{in} _N(R_d, \theta )$
may be used to evaluate the  coefficients $a^{'}_s$  in Eq.~(5)
allowing in this way to reconstruct the full wave function
$\Psi_N(r, \theta )$ of the billiard.

In the evaluation of the  coefficients  $a^{'}_s$ in Eq.~(5) an
important role plays the value of the refraction index $n$ of the
Teflon insert. We measured the refraction index $\eta =1.425 \pm
0.002$ of Teflon by measuring the set of resonant frequencies of a
microwave circular cavity of radius $R_T=3.25$ cm entirely filled
by it.

\begin{figure}[!]
\begin{center}
\rotatebox{270} {\includegraphics[width=0.5\textwidth,
height=0.6\textheight, keepaspectratio]{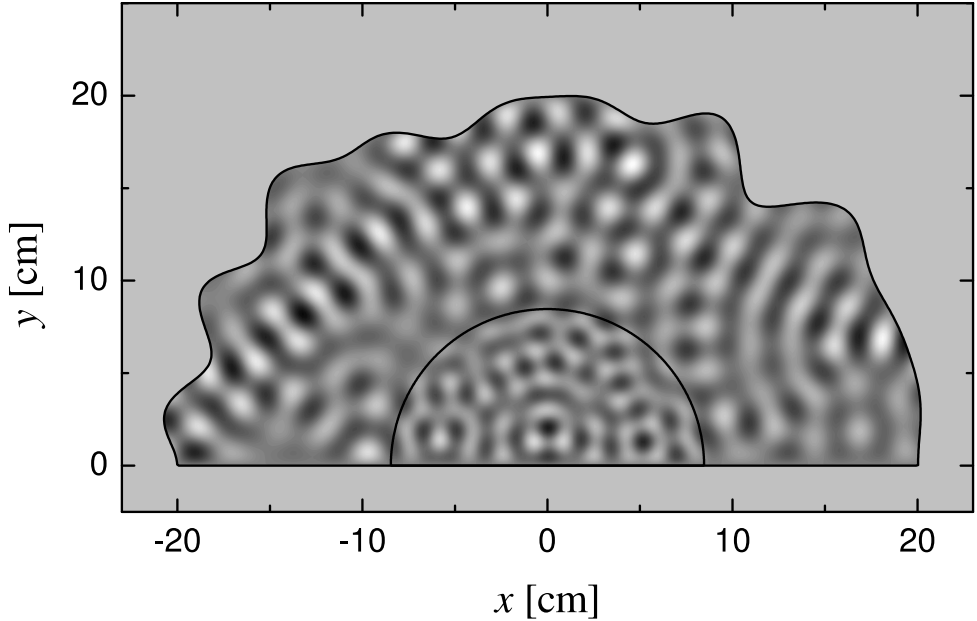}} \caption{ The
reconstructed wave function $\Psi_{415}(r,\theta ) $ of the
chaotic half-circular microwave rough billiard. The amplitudes
have been converted into a grey scale with white corresponding to
large positive and black corresponding to large negative values,
respectively. Dimensions of the billiard are given in cm. The
position of the half-circular Teflon insert of radius $R_d=8.465$
cm is marked with a solid line.}\label{Fig3}
\end{center}
\end{figure}

Using the method of the ``trial wave function" we were able to
reconstruct 30 experimental wave functions of the rough
half-circular billiard with the level number $N$ between 215 and
415. The wave functions were reconstructed on points of a square
grid of side $4.2 \cdot 10^{-4} $ m. As the quantitative measure
of the sign assignment quality we chose  the integral $\gamma
\int_{\Gamma }|\Psi_N(r,\theta )|^2dl $ calculated along the
billiard's rough boundary $\Gamma $, where $\gamma $ is length of
$\Gamma $. In Fig.~\ref{Fig2} (b) we show the ``trial wave
function" $\Psi_{415} (R_c, \theta )$ with the correctly assigned
signs, which was used in the  reconstruction of the wave function
$\Psi_{415}(r, \theta )$ of the billiard (see Fig.~\ref{Fig3}). It
is worth noting that inside of the Teflon insert the size of nodal
domains are much smaller than outside of it. The remaining wave
functions from the range $N=215-415$ were not reconstructed
because of the accidental near-degeneration of the neighboring
states or due to the problems with the measurements of
$|\Psi_N(R_c, \theta )|^2$ along a half-circle coinciding for its
significant part with one or several of the nodal lines of
$\Psi_N(r, \theta )$. The problem of the near-degenerated states
is important because in the presence of the perturber the
resonances are shifted, which may cause the initially
non-overlapping states to become near-degenerated at certain
positions of the perturber. Such a situation prevents us from the
reconstruction of the wave functions. The problems mentioned are
getting much more severe for $N>200$. Furthermore, the computation
time $t_r$ required for reconstruction of the "trial wave
function" scales like $t_r \propto 2^{n_z -2}$, where $n_z$ is the
number of identified zeros in the measured function $|\Psi_N(R_c,
\theta )|$.

\begin{figure}[!]
\begin{center}
\rotatebox{0} {\includegraphics[width=0.5\textwidth,
height=0.8\textheight, keepaspectratio]{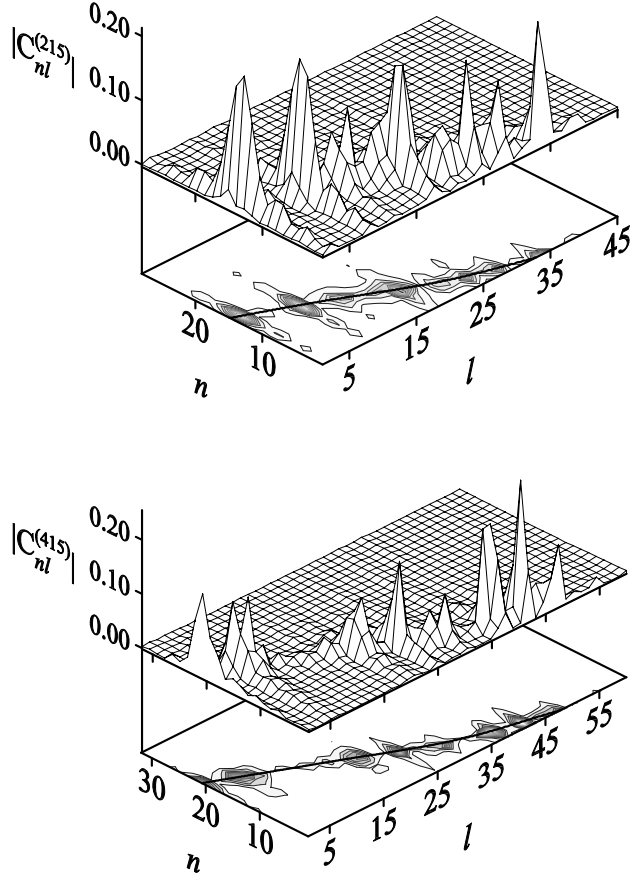}} \caption{Structure
of the energy surface in the regime of Shnirelman ergodicity. Here
we show the moduli of amplitudes $|C^{(N)}_{nl}|$ for the wave
functions: (a) $N=215$, (b) $N=415$. The wave functions are
delocalized in the $n, l$ basis. Full lines show the energy
surface (see text).}\label{Fig4}
\end{center}
\end{figure}

The structure of the energy surface \cite{Frahm97} of the
billiard's wave functions plays an important role in the
identification of their ergodicity. To check it we extracted wave
function amplitudes $C^{(N)}_{nl}=\left< n,l|N \right>$  in the
basis $n, l$ of a half-circular ray-splitting billiard
(desymmetrized annular ray-splitting billiard) \cite{Kohler1998}
with radius $r_{max}$ and a half-circular Teflon insert of radius
$R_d$ . The normalized eigenfunctions of the half-circular
ray-splitting billiard are given by
$$ \Phi_{nl} (r,\theta) = \left\{ \begin{array}{cc} A_{ln} J_{l}(\eta \kappa_{ln} r) \sin(l \theta ) ,
& 0 \leq r \leq R_{d}, \\
A_{ln} \left[ C_{ln} J_{l}(\kappa_{ln} r) + D_{ln}
Y_{l}(\kappa_{ln} r) \right]\sin(l \theta ) , & R_{d} \leq r \leq
r_{max},
\end{array}\right. \eqno(6)$$
where $ A_{ln} = \left\{ \frac{\pi}{2} \left( \int_0^{R_{d}} r
J_{l}(\eta \kappa_{ln} r) ^{2} dr + \int_{R_{d}}^{r_{max}} r
\left[ C_{ln} J_{l}(\kappa_{ln} r) + D_{ln} Y_{l} (\kappa_{ln} r)
\right]^{2} dr \right) \right\}^{-\frac{1}{2}} $.
$J_{l}(\kappa_{ln} r)$ and $Y_{l}(\kappa_{ln} r)$ are Bessel and
Neumann functions, respectively. The main quantum number $n=1,2,3
\ldots$ enumerates the zeros $y_{ln}=\kappa_{ln} r_{max}$ of the
radial function
$$ C_{ln} J_{l} (y_{ln}) + D_{ln} Y_{l} (y_{ln}) = 0 , \eqno(8)$$
and $ l=1,2,3 \ldots$ is the angular momentum quantum number. The
coefficients $C_{ln}$ and $D_{ln}$ can be determined from the
continuity conditions of the wave function $ \Phi_{nl} (r,\theta)$
and it's derivative $ \Phi_{nl}^{'} (r,\theta)$ on Teflon's
boundary $R_{d}$
$$ \left\{ \begin{array}{cc} J_{l}(\eta \kappa_{ln} R_{d}) = C_{ln} J_{l} (\kappa_{ln} R_{d}) +
D_{ln} Y_{l} (\kappa_{ln} R_{d}), \\ \eta J_{l}^{'} (\eta
\kappa_{ln} R_{d}) = C_{ln} J_{l}^{'}(\kappa_{ln} R_{d}) + D_{ln}
Y_{l} ^{'}(\kappa_{ln} R_{d}).
\end{array} \right.   \eqno(7)$$
The moduli of amplitudes $|C^{(N)}_{nl}|$ and their projections
into the energy surface for the representative experimental wave
functions $N=215$ and $N=415$ are shown in Fig.~\ref{Fig4}. As
expected, in the regime of Shnirelman ergodicity the wave
functions are extended over the whole energy surface
\cite{Hlushchuk01}. The full lines on the projection planes in
Fig.~\ref{Fig4}(a) and Fig.~\ref{Fig4}(b) mark the energy surface
of a half-circular annular ray-splitting billiard $H(n,l)\simeq
E_N=k^2_N$ estimated from the formula $ |H(n,l)-E_N|/E_N \leq
0.12$. The peaks $|C^{(N)}_{nl}|$ are spread almost perfectly
along the lines marking the energy surface.

\begin{figure}[!]
\begin{center}
\rotatebox{0} {\includegraphics[width=0.5\textwidth,
height=0.8\textheight, keepaspectratio]{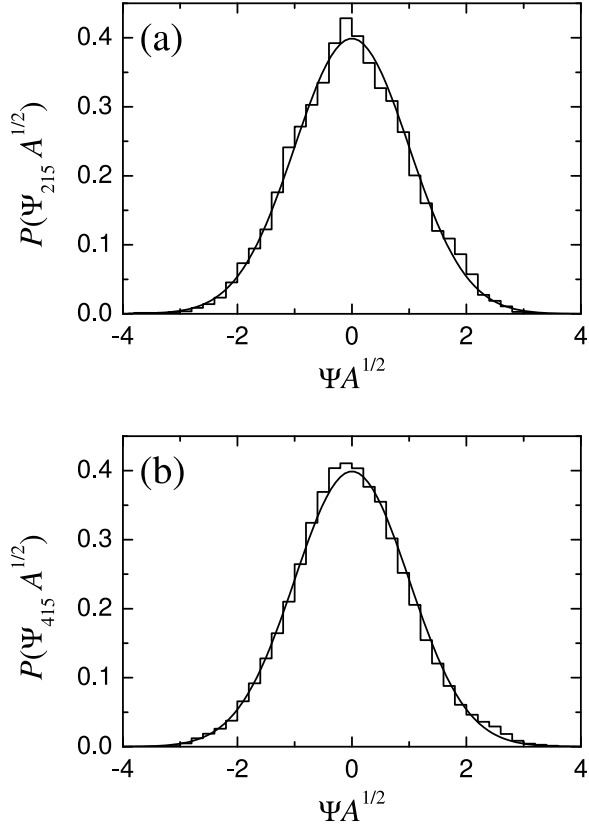}} \caption{Panel (a):
The amplitude distribution $P(\Psi_N A^{1/2})$ for the wave
function $N=215$. Panel (b): The distribution $P(\Psi_N A^{1/2})$
for the  wave function $N=415$. The amplitude distributions were
constructed as histograms with bin equal to 0.2. The width of the
distribution $P(\Psi)$ was rescaled to unity by multiplying
normalized to unity wave function by the factor $A^{1/2}$, where
$A$ denotes billiard's area. Full lines show standard normalized
Gaussian prediction $P_{0}(\Psi
A^{1/2})=(1/\sqrt{2\pi})e^{-\Psi^{2} A/2}$. }\label{Fig5}
\end{center}
\end{figure}

Ergodic behavior of the measured wave functions can be also tested
by evaluation of the amplitude distribution $P(\Psi_N)$
\cite{Berry77,Kaufman88}. For irregular, chaotic states the
probability of finding the value $\Psi_N$ at any point inside the
billiard should be distributed as a Gaussian, $P(\Psi_N) \sim
e^{-\beta \Psi_N^{2}}$. In Fig.~\ref{Fig5}(a) we show the
amplitude distribution $P(\Psi_N A^{1/2})$ for  the wave function
$N=215$  while in Fig.~\ref{Fig5}(b) the distribution $P(\Psi_N
A^{1/2})$ for the wave function $N=415$ is presented.  The
distributions were constructed as normalized to unity histograms
with the bin equal to 0.2. The width of the amplitude
distributions $P(\Psi_N )$ was rescaled to unity by multiplying
normalized to unity wave functions by the factor $A^{1/2}$, where
$A$ denotes billiard's area (see formula (23) in
\cite{Kaufman88}). For all measured wave functions lying in the
regime of Shnirelman ergodicity  the distributions of $P(\Psi_N
A^{1/2})$  were in good agreement with the standard normalized
Gaussian prediction $P_{0}(\Psi
A^{1/2})=(1/\sqrt{2\pi})e^{-\Psi^{2} A/2}$.

\begin{figure}[!]
\begin{center}
\rotatebox{270} {\includegraphics[width=0.5\textwidth,
height=0.6\textheight, keepaspectratio]{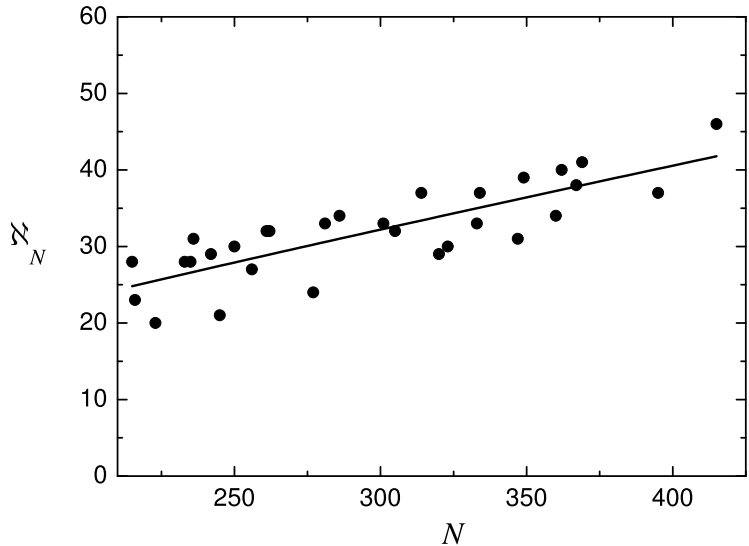}} \caption{The number
of nodal domains $\aleph_N$ (full circles) for the chaotic
half-circular microwave ray-splitting rough billiard. Full line
shows the least squares fit $\aleph_N = a_1N +b_1\sqrt{N}$ to the
experimental data (see text), where $a_1=0.063 \pm 0.023$,
$b_1=0.77 \pm 0.40$. The prediction of the theory of Bogomolny and
Schmit \cite{Bogomolny2002} $a_1=0.062$. } \label{Fig6}
\end{center}
\end{figure}

The number of nodal domains $\aleph_N$ vs. the level number $N$ in
the chaotic microwave ray-splitting rough billiard is plotted in
Fig.~\ref{Fig6}. The full line in Fig.~\ref{Fig6} shows the least
squares fit $\aleph_N = a_1N +b_1\sqrt{N}$ of the experimental
data, where $a_1=0.063 \pm 0.023$, $b_1=0.77 \pm 0.40$. The
coefficient $a_1=0.063 \pm 0.023$ coincides with the prediction of
the percolation model of Bogomolny and Schmit \cite{Bogomolny2002}
$\aleph_N/N \simeq 0.062$ within the error limits. The errors of
the coefficients $a_1$ and $b_1$ are relatively high because the
number of nodal domains fluctuates significantly in the function
of the level number $N$, what was also demonstrated in
\cite{Blum2002} (see Fig~.(5)). It is worth mention that in the
paper \cite{Savytskyy2004} the coefficient $a_1$ was estimated in
the experiment with the microwave rough billiard without the
ray-splitting Teflon insert. Its value $a_1=0.058 \pm 0.006$ was
also close to the theoretical prediction. The second term in the
least squares fit corresponds to a contribution of boundary
domains, i.e. domains that include the billiard boundary.
Numerical calculations of Blum {\it et al.} \cite{Blum2002}
performed for the Sinai and stadium billiards showed that the
number of boundary domains scales as the number of the boundary
intersections, that is as $\sqrt{N}$. Present results together
with the results of \cite{Savytskyy2004} clearly suggest that in
the rough billiards (with and without ray-splitting), at low level
number $N$, the boundary domains also significantly influence the
scaling of the number of nodal domains $\aleph_N$, leading to the
departure from the predicted scaling $\aleph_N \sim N$.

\begin{figure}[!]
\begin{center}
\rotatebox{270} {\includegraphics[width=0.5\textwidth,
height=0.6\textheight, keepaspectratio]{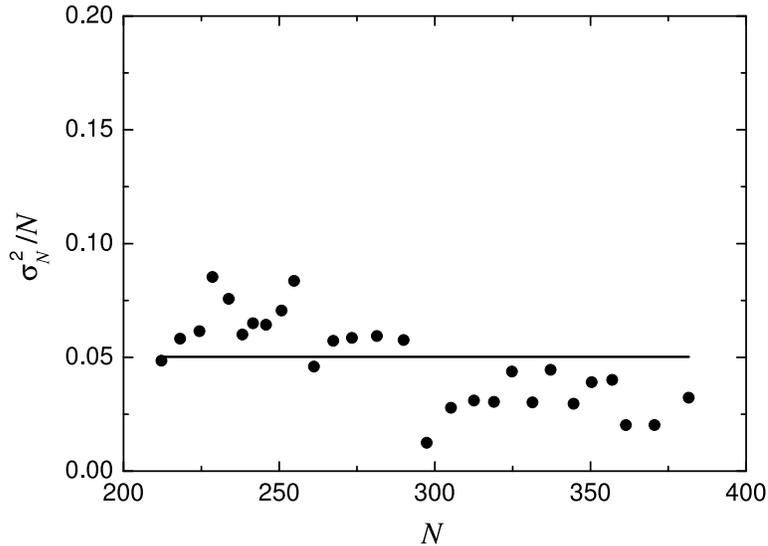}} \caption{The
variance of the mean number of nodal domains divided by the level
number $\sigma^2_N /N$ for the chaotic half-circular microwave
ray-splitting rough billiard. Full line shows predicted by the
theory limit  $\sigma^2_N /N \simeq 0.05 $, Bogomolny and Schmit
\cite{Bogomolny2002}.} \label{Fig7}
\end{center}
\end{figure}

Measured wave functions of the ray-splitting billiard may be also
used for the calculations of the variance $\sigma^2_N $ of the
mean number of nodal domains. It was predicted in
\cite{Bogomolny2002} that for chaotic wave functions the variance
of the mean number of nodal domains should converge to the
theoretical limit $\sigma^2_N \simeq 0.05N$. In Fig.~\ref{Fig7}
the variance of the mean number of nodal domains divided by the
level number $\sigma^2_N /N$ is shown for the microwave
ray-splitting rough billiard.  The variance $\sigma^2_N =
\frac{1}{N_w-1}\sum_{i=1}^{N_w} (\aleph_{N_i}
-\langle\aleph_N\rangle)^2$ was calculated in the window  of
$N_w=5$ consecutive eigenstates measured between $215\leq N \leq
415$, where the mean number of nodal domains was defined as
$\langle\aleph_N\rangle = \frac{1}{N_w}\sum_{i=1}^{N_w}
\aleph_{N_i} $. For level numbers $N < 300$ the variance
$\sigma^2_N /N$ is above the predicted theoretical limit, however,
for $300<N\leq 415$ it is slightly below it. A similar erratic
behavior  of $\sigma^2_N /N$ was also observed in
\cite{Bogomolny2002}.

\begin{figure}[!]
\begin{center}
\rotatebox{270} {\includegraphics[width=0.5\textwidth,
height=0.6\textheight, keepaspectratio]{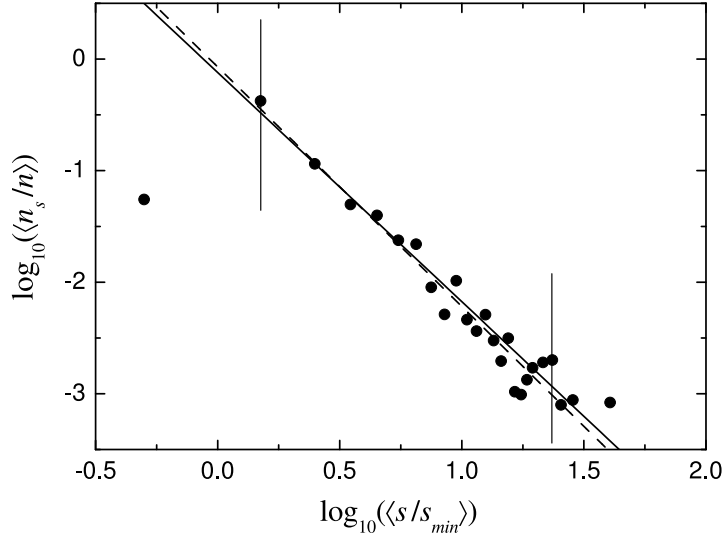}}
\caption{Distribution of nodal domain areas. Full line shows the
prediction of percolation theory $\log_{10}(\langle n_s/n \rangle)
= -\frac{187}{91} \log_{10}(\langle s/s_{min} \rangle)$. The least
squares fit $\log_{10}(\langle n_s/n \rangle) = a_2 -\tau
\log_{10}(\langle s/s_{min} \rangle)$ of the experimental results
lying within the vertical lines yields the scaling exponent $\tau
= 2.14 \pm 0.12$ and $a_2=-0.06 \pm 0.12$.  The result of the fit
is shown by the dashed line. }\label{Fig8}
\end{center}
\end{figure}

The percolation model \cite{Bogomolny2002} allows for applying the
results of percolation theory to the description of nodal domains
of chaotic billiards. The percolation theory predicts that the
distribution of the areas $s$  of nodal clusters should obey the
scaling behavior:  $n_s \propto s^{-\tau}$. The scaling exponent
\cite{Ziff1986} is found to be $\tau = 187/91$. In Fig.~\ref{Fig8}
we present in logarithmic scales nodal domain areas distribution
$\langle n_s/n \rangle$ vs. $\langle s/s_{min} \rangle$ obtained
for the microwave ray-splitting rough billiard. The distribution
$\langle n_s/n \rangle$ was constructed as normalized to unity
histogram with the bin equal to 1.  The areas $s$ of nodal domains
were calculated by summing up the areas of the nearest neighboring
grid sites having the same sign of the wave function. In
Fig.~\ref{Fig8} the vertical axis $\langle n_s/n \rangle =
\frac{1}{N_T}\sum_{i=1}^{N_T}n_s^{(N)}/n^{(N)} $ represents the
number of nodal domains $n_s^{(N)}$ of size $s$ divided by the
total number of domains $n^{(N)}$ averaged over $N_T=30$ wave
functions measured in the range $215 \leq N \leq 415$. In these
calculations we took into account only the nodal domains which
entirely lied outside or inside of the Teflon insert for which
percolation theory \cite{Ziff1986} should be applicable. The
horizontal axis in Fig.~\ref{Fig8} is expressed in the units of
the smallest possible area $s_{min}^{(N)}$ \cite{Bogomolny2002},
$\langle s /s_{min} \rangle =
\frac{1}{N_T}\sum_{i=1}^{N_T}s/s_{min}^{(N)} $, where
$s_{min}^{(N)}=\pi(j_{01}/\eta k_N)^2$ and $j_{01}\simeq 2.4048$
is the first zero of the Bessel function $J_0(j_{01})=0$. For
nodal domains lying inside the Teflon insert the refraction index
was according to our measurements $\eta =1.425$ while outside of
the insert we assumed $\eta =1$. The full line in Fig.~\ref{Fig8}
shows the prediction of percolation theory $\log_{10}(\langle
n_s/n \rangle) = -\frac{187}{91} \log_{10}(\langle s/s_{min}
\rangle)$. In a broad range of $\log_{10}(\langle s/s_{min}
\rangle)$, approximately from 0.2 to 1.4, which is marked by the
two vertical lines the experimental results follow closely the
theoretical prediction. The least squares fit $\log_{10}(\langle
n_s/n \rangle) = a_2 -\tau \log_{10}(\langle s/s_{min} \rangle)$
of the experimental results lying within the vertical lines gives
the scaling exponent $\tau = 2.14 \pm 0.12$ and $a_2=-0.06 \pm
0.12$, which is in a good agreement with the predicted $\tau=
187/91 \simeq 2.05$.  The result of the fit is shown in
Fig.~\ref{Fig8} by the dashed line.

In summary, for the first time we measured high-lying wave
functions of the chaotic microwave ray-splitting rough billiard.
We showed that in the limit $N \rightarrow \infty $ the least
squares fit of the experimental data reveals the asymptotic number
of nodal domains $\aleph_N/N \simeq 0.063 \pm 0.023$ that is close
to the theoretical prediction $\aleph_N/N \simeq 0.062$
\cite{Bogomolny2002}. We demonstrate that for higher level numbers
$N \simeq 215-415$ the variance of the mean number of nodal
domains $\sigma^2 /N$ is scattered around the theoretical limit
$\sigma^2 /N \simeq 0.05$. Following the results of
percolationlike model proposed by \cite{Bogomolny2002} we
confirmed that the distribution of the areas $s$  of nodal domains
has power behavior $n_s \propto s^{-\tau}$, where scaling exponent
is equal to $\tau = 2.14 \pm 0.12$. This result is in a good
agreement with the prediction of percolation theory
\cite{Ziff1986}, which predicts $\tau= 187/91 \simeq 2.05$. The
experimental results presented in this paper strongly suggest that
many properties of nodal domains in chaotic ray-splitting
billiards are the same, like in  conventional chaotic  billiards
without ray-splitting.

Acknowledgments.  This work was supported by Ministry of Science
and Information Society Technologies grant No. 2 P03B 047 24.

\end{document}